\documentclass[
  twocolumn,
  showpacs,
  amsmath,
  amssymb,
  prl
]{revtex4}

\usepackage{bm}      
\usepackage{amsthm}  

\newcommand*{\myBoldsymbol}[1]{{\bm{#1}}}  
\newcommand*{\myEnsuremath}[1]{#1}  
\newcommand*{\myHat}[1]{\Hat{#1}}
\newcommand*{\myBar}[1]{\Bar{#1}}
\newcommand*{\myMathbb}[1]{\mathbb{#1}}
\newcommand*{\myDot}[1]{{\dot{#1}}}
\newcommand*{\myDdot}[1]{{\ddot{#1}}}
\newcommand*{\sbar}[1]{\myEnsuremath{\myBar{#1}}}

\newcommand*{\sddot}[1]{\myEnsuremath{\myDdot{#1}}}
\newcommand*{\gsb}[1]{\myEnsuremath{\myBoldsymbol{#1}}}
\newcommand*{\gsbbar}[1]{\myBar{\gsb{#1}}}

\newcommand*{\lb}[1]{\gsb{#1}}
\newcommand*{\lbhat}[1]{\myHat{\lb{#1}}}
\newcommand*{\lbdot}[1]{\myDot{\lb{#1}}}

\newcommand*{\lbb}[1]{\myEnsuremath{\myMathbb{#1}}}
\newcommand*{\braket}[3]{\myEnsuremath{\langle#1|#2|#3\rangle}}
\newcommand*{\inner}[2]{\myEnsuremath{\langle#1|#2\rangle}}
\newcommand*{\ket}[1]{\myEnsuremath{|#1\rangle}}
\newcommand*{\bra}[1]{\myEnsuremath{\langle#1|}}

\newcommand{\myeq}[1]{(\ref{#1})}
\newcommand{\myonehalf}{\tfrac{1}{2}}

\DeclareMathOperator{\mytr}{tr}      

\DeclareMathOperator{\mysgn}{sgn}

\theoremstyle{definition}            
\newtheorem{theorem}{Theorem}
\newtheorem{lemma}{Lemma}
\newtheorem{problem}{Problem}

\begin{document}
\title{Positive $P$-Representations of the Thermal Operator\\
from Quantum Control Theory}
\author{John A. Sidles} 
\email[Email:]{sidles@u.washington.edu} 
\homepage[\\ 
URL: ]{http://courses.washington.edu/goodall/MRFM/}
\thanks{\\ 
Supported by the NIH/NCRR, the NSF/ECS, the Army
Research Office (ARO), and DARPA/DSO/MOSAIC.}
\affiliation{Quantum System Engineering Group\\
University of Washington, School of Medicine\\
Box 356500, Seattle, WA 98195, USA}
\date{January 26, 2004}
\begin{abstract}
A positive $P$-representation for the spin-$j$ thermal
density matrix is given in closed form. The representation
is constructed by regarding the wave function as the
internal state of a closed-loop control system. A
continuous interferometric measurement process is proved
to einselect coherent states, and feedback control is
proved to be equivalent to a thermal reservoir. Ito
equations are derived, and the $P$-representation is
obtained from a Fokker-Planck equation. Langevin equations
are derived, and the force noise is shown to be the
Hilbert transform of the measurement noise. The formalism
is applied to magnetic resonance force microscopy (MRFM)
and gravity wave (GW) interferometry. Some unsolved
problems relating to drift and diffusion on Hilbert
spaces are noted.
\end{abstract}

\pacs{42.50.Lc, 02.30.Yy, 03.65.Yz, 05.40.-a}

\maketitle

Given spin-$j$ operators 
$\lb{s}= (s_{1}, s_{2},s_{3})$,
it is well known
\cite{Perelomov:72,Perelomov:86,Gardiner:00,Radcliffe:70}
that a coherent state $\ket{\lbhat{x}}$ can be associated
with each unit vector $\lbhat{x}$, such that
$\inner{\lbhat{x}}{\lbhat{x}} = 1$,
$\braket{\lbhat{x}}{\lb{s}}{\lbhat{x}}= j \lbhat{x}$, and
a resolution of the identity operator $\lbb{I}$ is
\begin{equation}
  \lbb{I} = \frac{2 j+1}{4\pi}\int_{4\pi\negthickspace}
  \!\!\,d^{2}\lbhat{x}\,\ket{\lbhat{x}}\bra{\lbhat{x}}\,.
\end{equation}

Given an operator $\rho_{j}$ on each member of a sequence
of spin-$j$ Hilbert spaces $j\in
\{0,\myonehalf,1,\ldots\}$, we define a
\emph{$P$-sequence} to be a set of functions
$\{P_{j}(\lbhat{x})\}$, and similarly a
\emph{$Q$-sequence} to be a set of functions
$\{Q_{j}(\lbhat{x})\}$, satisfying
\begin{subequations}
\begin{align}
  \rho_{j} & = \frac{2j+1}{4\pi}\int_{4\pi\negthickspace}
  \!\!\,d^{2}\lbhat{x}\,P_{j}(\lbhat{x})\,
  \ket{\lbhat{x}}\bra{\lbhat{x}}
  \,,
  \label{eq:Prep}\\
   \braket{\lbhat{x}}{\rho_{j}}{\lbhat{x}}& 
   =Q_{j}(\lbhat{x})\,.
  \label{eq:Qrep}
\end{align}
\end{subequations}
The properties of $P$-sequences and $Q$-sequences will be
the main topic of this article.

$Q$-sequences pose few mathematical mysteries because
\myeq{eq:Qrep} suffices to evaluate $Q_{j}(\lbhat{x})$
directly. In
contrast, little is known about $P_{j}$-representations,
particularly about their positivity properties. The main
obstruction is that $P_{j}$ is nonunique
due to the identity \cite{Perelomov:72,Perelomov:86} 
\begin{alignat}{2}
  \int_{4\pi\negthickspace}\!\!\,d^{2}\lbhat{x}\ 
  Y^{l}_{m}(\lbhat{x})\,\ket{\lbhat{x}}\bra{\lbhat{x}}
  &= 0 &\quad
  \text{for $l>2j$,}
   \label{eq:indeterminacy}
\end{alignat}
where $Y^{l}_{m}(\lbhat{x})$ is a spherical harmonic. To
circumvent this $P$-representation ambiguity we will apply
methods from quantum measurement and control theory.

We will study the $P$-sequence of the thermal operator.
\begin{equation}
\label{eq:thermal operator}
  \rho^{\text{th}} = 
  \exp(-\beta \lbhat{t}\gsb{\cdot}\lb{s})\,,
\end{equation}
where $\lbhat{t}$ is the thermal axis. The $Q$-sequence of
$\rho^{\text{th}}$ can be computed by direct evaluation of
\myeq{eq:Qrep}, and the well-known result
\cite{Radcliffe:70,Perelomov:86} is
\begin{equation}
  Q_{j}(\lbhat{x}) = (\cosh\myonehalf\beta -
  \lbhat{x}\gsb{\cdot}\lbhat{t}
  \sinh\myonehalf\beta)^{2j}\,.
\end{equation}
Our main result is:
\begin{subequations}
\begin{theorem} 
\label{th:theorem1}
\emph{A positive $P$-sequence for the thermal operator is} 
$P_{j}(\lbhat{x})  =c_{j}/Q_{j+1}(-\lbhat{x})$.
\end{theorem}
\noindent The normalization coefficient $c_{j}$ is readily
found from \myeq{eq:Prep} by taking the trace, with the
result
\begin{equation}
  \label{eq:theorem2}
  c_{j} = \frac{2\sinh\gsb{(}(j+\myonehalf)\beta\gsb{)}}{
  Q^{-1}_{j+1/2}(\lbhat{t})-
  Q^{-1}_{j+1/2}(-\lbhat{t})}\,.
\end{equation}
\end{subequations}

We first prove Theorem~\ref{th:theorem1} nonconstructively
by treating it as an \emph{ansatz}. We take the thermal
axis $\lbhat{t}=(0,0,1)$ such that
$\braket{j,m}{\rho^{\text{th}}}{j,m'}=\delta_{mm'}e^{-\beta
m}$. Taking matrix elements of \myeq{eq:Prep},
Theorem~\ref{th:theorem1} is proved if $P_{j}$ satisfies
\begin{equation}
  e^{-\beta m}\delta_{mm'}
  = \frac{2
  j+1}{4\pi}\int_{4\pi\negthickspace}
  \!\!\,d^{2}\lbhat{x}\,P_{j}(\lbhat{x})\,
  \inner{j,m}{\lbhat{x}}\inner{\lbhat{x}}{j,m'}\,.
  \label{eq:mmprime}
\end{equation}
We write $\ket{\lbhat{x}} = D(\phi,\theta,0)\ket{j,j}$,
where the rotation
\begin{equation}
  D(\phi,\theta,\psi)=e^{-i\phi s_{3}} e^{-i\theta s_{2}}
  e^{-\psi s_{3}}
\end{equation}
carries $\lbhat{t}=(0,0,1)$ into $\lbhat{x}=(\sin\theta
\cos\phi, \sin\theta \sin\phi, \cos\theta)$. By an identity
due to Wigner \cite{Gottfried:66,Rose:57}
\begin{multline}
  \inner{j,m}{\lbhat{x}} = 
  \braket{j,m}{D(\phi,\theta,0)}{j,j} = D^{j}_{mj}(\phi,\theta,0) \\
  = \tbinom{2j}{j-m}^{1/2}
  e^{-i m\phi}
  \cos^{j+m}\!\myonehalf\theta\ 
  \sin^{j-m}\!\myonehalf\theta\,,
\end{multline}
we transform \myeq{eq:mmprime} into an integral
representation of the hypergeometric function
$\sb{2}F_{1}$. Then via the identity \cite{Abramowitz:65}
\begin{equation}
  \label{eq:hypergeometric}
  \sb{2}F_{1}(2 + 2 j, 1 + j + m, 2 + 2 j, z)
= (1-z)^{-(j+m+1)}.
\end{equation}
Theorem~\ref{th:theorem1} follows immediately. 

But where did the \emph{ansatz} come from? Why do positive
$P$-representations even exist for thermal operators?

We will now give a constructive proof that answers these
questions. Our notation and methods are adapted from
\emph{Handbook of Stochastic Processes}
\cite{Gardiner:85}. Purely algebraic details are not
given, but when non-obvious idioms or strategies are
employed, we will show them.

The proof consists of three physically-motivated steps. In
the first step, we will regard $\ket{\psi}$ as the
internal state of an open-loop dynamical system, and we
will specify a sensor that monitors the spin axis.

In the second step, we will install a controller that uses
the (imperfect) sensor measurements to align the spin axis
with the thermal axis $\lb{t}$. We will prove that the
control noise is precisely equivalent to a thermal bath.

In the third and final step, we will construct Ito and
Fokker-Planck equations for the observed states.
Theorem~\ref{th:theorem1} then emerges constructively,
with $P_{j}$ as the solution of a Fokker-Planck equation.

To begin Step~1, we define an \emph{open-loop uniaxial
spinometer} with \emph{generator} $s$ to be the Markov
chain of quantum states (see, \emph{e.g.},
\cite{Gardiner:00}) defined by
\begin{equation}
  \label{eq:stochastic}
  \ket{\psi_{n+1}} = 
  \begin{cases}
    A\ket{\psi_{n}}/\sqrt{P_\text{A}};&P_{\text{A}} = 
    \braket{\psi_{n}}{A^{\dagger}A}{\psi_{n}}\\
    B\ket{\psi_{n}}/\sqrt{P_\text{B}};&P_{\text{B}} = 
    \braket{\psi_{n}}{B^{\dagger}B}{\psi_{n}}
  \end{cases}
\end{equation}
with increment operators $A$ and $B$ given by
\begin{subequations}
\begin{align}
  \label{eq:stochasticA}
    A &= [\cos (\theta s) + \sin (\theta s)]/\sqrt{2}\,,\\
    B &= [\cos (\theta s) - \sin (\theta s)]/\sqrt{2}\,.
    \label{eq:stochasticB}
\end{align}
\end{subequations}
The generator $s$ can be any Hermitian operator. 

Here $P_{\text{A}}+P_{\text{B}}=1$ is guaranteed by
$A^{\dagger}A+B^{\dagger}B = \lbb{I}$. The measurement
strength is governed by $\theta$, and we will assume
$\theta\ll 1$. Terms of order $\theta^{3}$ and higher will
turn out to be nonleading and we will ignore them.

We define the operator variance $\Delta(s)$ to be
\begin{equation}
  \label{eq:variance}
  \Delta_{n}(s) = \braket{\psi_{n}}{s^{2}}{\psi_{n}}-
  \braket{\psi_{n}}{s}{\psi_{n}}^{2}\,.
\end{equation}
It is readily shown that in general 
\begin{equation}
\label{eq:uniaxial inequality}
\Delta_{n}(s)
\begin{cases}
  =0&\text{if $\ket{\psi_{n}}$ is an eigenstate of $s$,}\\
  >0& \text{otherwise}.
\end{cases}
\end{equation}
For $E$ an ensemble average, the variance increment
\begin{equation}
  \label{eq:deltaDef}
\delta E_{n}[\Delta(s)]  =
E[\Delta_{n+1}(s)\lb]-E[\Delta_{n}(s)]
\end{equation}
is found by straightforward stochastic analysis to be
\begin{equation}
\label{eq:theoremI}
\delta E_{n}[\Delta(s)]   =
  -4\theta^{2} E[\Delta^{2}_{n}(s)]\,.
\end{equation}
The sequence
$\{E[\Delta_{1}(s)],E[\Delta_{2}(s)],\ldots\}$ is positive
by \myeq{eq:uniaxial inequality} and decreasing by
\myeq{eq:theoremI}, therefore it has a limit. This implies
a vanishing increment \myeq{eq:theoremI}, which implies
$\lim_{n\to\infty} E_{n}[\Delta^{2}(s)]=0$, which implies
$\lim_{n\to\infty} \Delta_{n}(s)=0$ for every Markov chain
in the ensemble (except a set of measure zero). This
proves
\begin{theorem}
  \label{th:uniaxial}
  \emph{Open-loop uniaxial spinometers asymptotically
  einselect eigenstates of the generator}.
\end{theorem}
\noindent Here we have used \emph{einselect} in Zurek's sense
\cite{Zurek:03}: 
\begin{quote}
  \emph{%
  Einselection [excludes] all but a
  small set [of states] from within a much larger Hilbert
  space. Einselected states are distinguished by their
  stability in spite of the monitoring environment.}
\end{quote}
\noindent Now we operate three spinometers simultaneously,
with generators $(s_{1},s_{2},s_{3})$. We call this an
\emph{open-loop triaxial spinometer}. We define a spin
covariance $\gsb{\sigma}_{n}$
\begin{equation}
(\gsb{\sigma}_{n})_{kl} = 
\braket{\psi_{n}}{s_{k}s_{l}}{\psi_{n}}-
\braket{\psi_{n}}{s_{k}}{\psi_{n}}
  \braket{\psi_{n}}{s_{l}}{\psi_{n}}
\end{equation}
and similar to \myeq{eq:variance} and \myeq{eq:deltaDef}
we define an $n$'th stochastic increment for the $i'th$
spinometer
\begin{equation} 
\delta E_{i,n}[\gsb{\sigma}] =
E[\gsb{\sigma}_{n+1}]|_{s=s_{i}}-E[\gsb{\sigma}_{n}]\,.
\end{equation}
By straightforward stochastic analysis similar to
\myeq{eq:theoremI}, the net covariance increment is
calculated to be
\begin{equation}
  \label{eq:triaxial increment}
\mytr\delta E_{n}[\gsb{\sigma}]  =  \sum_{i=1}^{3}\mytr\delta
E_{i,n}[\gsb{\sigma}] 
  = -4\theta^{2}
  \mytr E[\gsb{\sigma}_{n}\gsb{\cdot}\gsb{\sigma}_{n}^{\star}]\,.
\end{equation}
Physical intuition then suggests that similar to
\myeq{eq:uniaxial inequality}
\begin{equation}
\label{eq:asymmetric inequality}
\mytr\gsb{\sigma}_{n}\gsb{\cdot}\gsb{\sigma}_{n}^{\star}
\begin{cases}
  =0&\text{if and only if $\ket{\psi_{n}}$ is coherent,}\\
  >0& \text{otherwise.}
\end{cases}
\end{equation}
To prove this we define a spin vector $\lb{x} =
\braket{\psi}{\lb{s}}{\psi}/j$, such that
$\sigma_{lm}-\sigma_{ml} = i\,\epsilon_{lmn}\,jx_{n}$.
Without loss of generality we choose a frame in which
$\lb{x}=(0,0,x_{3})$. We define the symmetric tensor
$\sbar{\sigma}_{lm} = \myonehalf(\sigma_{lm} +
\sigma_{ml})$, and we temporarily regard $x_{3}$ and
$\{\sbar{\sigma}_{kl};k\ge l\}$ as a set of seven
arbitrary real numbers. Then solely by reordering terms we
construct two algebraic equalities
\begin{subequations}
\newcommand{\shortplus}{\!+\!}
\begin{align}
\label{eq:asymmetric decomposition}
\mytr \gsb{\sigma}\gsb{\cdot}\gsb{\sigma}^{\star}
& = p_{\text{a}}\shortplus
2 p_{\text{b}}\shortplus
\tfrac{1}{2}p_{\text{c}}\shortplus
j p_{\text{d}}\shortplus
\tfrac{1}{2}p_{\text{d}}^{2}\shortplus
\tfrac{1}{2}p_{\text{e}}\shortplus
\tfrac{1}{2}p_{\text{f}}\,\\
\mytr \gsbbar{\sigma}\gsb{\cdot}\gsbbar{\sigma}
\phantom{{}^{\star}}
& = \myonehalf j^{2}\shortplus
p_{\text{a}}\shortplus
2 p_{\text{b}}\shortplus
\tfrac{1}{2}p_{\text{c}}\shortplus
jp_{\text{d}}\shortplus
\tfrac{1}{2}p_{\text{d}}^{2}\shortplus
\tfrac{1}{2}p_{\text{f}}\,
\label{eq:symmetric decomposition}
\end{align}
\end{subequations}
whose individual terms are
\begin{alignat*}{2}
  p_{\text{a}} &= (\sbar\sigma_{33}-x^{2}_{3})^{2} &\qquad
  p_{\text{e}} &= j^{2}-\sbar{\sigma}_{33}\\
  p_{\text{b}} &=  \sbar{\sigma}_{12}^{2}+
  \sbar{\sigma}_{13}^{2}+\sbar{\sigma}_{23}^{2} &
  p_{\text{f}} &= j^{2}(1-x_{3}^{2})\\
  p_{\text{c}} &= (\sbar{\sigma}_{11}-\sbar{\sigma}_{22})^{2}& &\\
  p_{\text{d}} & = 
  (\mytr \gsbbar{\sigma}-\sbar{\sigma}_{33})^{2}-
  (j(j+1)-\sbar{\sigma}_{33})^{2}\,. 
  \hspace*{-12em}& &
\end{alignat*}
By construction, each term is nonnegative for all $\ket{\psi}$
and zero for coherent $\ket{\psi}$. In particular,
$p_{\text{e}}$ and $p_{\text{f}}$ vanish if and only if
$\sbar\sigma_{33}=\braket{\psi}{s_{3}^{2}}{\psi}=j^{2}$
and $jx_{3}=\braket{\psi}{s_{3}}{\psi}=\pm j$,
\emph{i.e.}, if and only if $\ket\psi$ is coherent. Then
\myeq{eq:asymmetric inequality} follows immediately from
the nonnegativity of \myeq{eq:asymmetric decomposition}.
Similarly \myeq{eq:symmetric decomposition} implies a
related inequality for $\gsbbar{\sigma}$:
\begin{equation}
\label{eq:symmetric inequality}
\mytr\gsbbar{\sigma}\gsb{\cdot}\gsbbar{\sigma}
\begin{cases}
  = \myonehalf j^{2}&\text{if and only if $\ket{\psi}$ is coherent,}\\
  >\myonehalf j^{2}& \text{otherwise.}
\end{cases}
\end{equation}

Then by the same reasoning as for
Theorem~\ref{th:uniaxial}, \myeq{eq:triaxial increment}
implies that $\lim_{n\to\infty} \mytr
\gsb{\sigma}_{n}\gsb{\cdot}\gsb{\sigma}_{n}^{\star}=0$ for
every chain in the ensemble (except a set of measure
zero). This proves
\begin{theorem}
  \label{th:triaxial}
  \emph{Open-loop triaxial spinometers asymptotically
  einselect coherent states}.
\end{theorem}
\noindent This completes Step~1 of our program. 

Remark: it seems reasonable that \emph{all} Lie groups
might asymptotically einselect coherent states
\cite{Perelomov:72}, but the author has studied only
$SU(2)$ $P$-representations.

Now we begin Step~2, and focus on control and
thermodynamics. For $\lbhat{t}$ the thermal axis defined
in \myeq{eq:thermal operator}, we modify the spinometer
matrices \myeq{eq:stochasticB} such that
\begin{subequations}
\begin{align}
  \label{eq:control}
    A^{\text{c}}_{k} &= 
    e^{-i \alpha (\lbhat{t}\times\lb{s})_{k}}
    [\cos (\theta s_{k}) + \sin (\theta s_{k})]/\sqrt{2}\,,\\
    B^{\text{c}}_{k} &= 
    e^{+i \alpha (\lbhat{t}\times\lb{s})_{k}}
    [\cos (\theta s_{k}) - \sin (\theta s_{k})]/\sqrt{2}\,.
\end{align}
\end{subequations}
We will call this a \emph{closed-loop triaxial spinometer}
with \emph{unitary feedback}, because the 
operators $\exp (\pm i \alpha\,\lbhat{t}\times\lb{s})$ act
cumulatively to align the spin axis with $\lbhat{t}$.

Closing the control loop does not alter the coherent
einselection because the sole effect of a \emph{post hoc}
unitary operator on $\gsb\sigma_{n}$ is a spatial
rotation. Since 
$\mytr\gsb{\sigma}_{n}\gsb{\cdot}\gsb{\sigma}_{n}^{\star}$ 
is a scalar, \myeq{eq:triaxial increment} still holds.
Thus we have
\begin{lemma}
  \label{th:triaxial feedback}
  \emph{Closed-loop triaxial spinometers with unitary
  feedback asymptotically einselect coherent states}.
\end{lemma}

An ensemble of closed-loop spinometers has a density
matrix sequence $\{\rho_{1},\rho_{2}, \ldots\}$ whose
increment is
\begin{equation}
  \label{eq:thermal increment}
  \delta\rho_{n} =\sum_{k=1}^{3} \left(
  A^{\text{c}\dagger}_{k}\rho_{n}A^{\text{c}}_{k} +
  B^{\text{c}\dagger}_{k}\rho_{n}B^{\text{c}}_{k}
  -\rho_{n}\right)\,,
\end{equation}
and we readily compute that $\delta\rho_{n}=0$ for
$\rho_{n}=\rho^{\text{th}}$, with $\rho^{\text{th}}$ the
thermal operator defined in \myeq{eq:thermal operator},
provided
\begin{equation}
  \label{eq:alpha beta}
  \alpha = -\tanh \tfrac{1}{4}\beta
  \quad\text{or}\quad
  1/\alpha = -\tanh \tfrac{1}{4}\beta\,.
\end{equation}
We will show later on that $\rho^{\text{th}}$ solves
$\delta\rho_{n}=0$ uniquely, because the Fokker-Planck
equation for $\rho$ has a unique stationary solution. This
proves
\begin{theorem}
  \label{th:thermal theorem}
  \emph{The density matrix of an ensemble of closed-loop
  triaxial spinometers with unitary feedback is
  asymptotically thermal.}
\end{theorem}

To connect \myeq{eq:thermal increment} with the
thermodynamic literature, we set $\lbhat{t}=(0,0,1)$ and
expand to order $\theta^{2}$. The result is equivalent to
a thermal model given by Perelomov (per eq.~23.2.1
of~\cite{Perelomov:86}). Gardiner gives similar model,
(per eq.~10.4.2 of \cite{Gardiner:85}). In Lindblad form
we find
\begin{equation}
  \label{eq:Lindblad}
  \begin{array}[b]{r@{\,}l}
  \delta\rho_{n} = 
  -\tfrac{1}{2} \gamma(\nu+1)&(s_{+}s_{-}\rho-2s_{-}\rho s_{+}+s_{+}s_{-})\\
  -\ \tfrac{1}{2}\gamma\nu&(s_{-}s_{+}\rho-2s_{+}\rho s_{-}+s_{-}s_{+})\\
  +\ \theta^{2}&(s_{3}s_{3}\rho-2s_{3}\rho s_{3}+s_{3}s_{3})\,,
  \end{array}
\end{equation}
where $s_{+}= (s_{1}+ i s_{2})/\sqrt{2}$ and 
$s_{-}= (s_{1} - i s_{2})/\sqrt{2}$, 
and we have adopted Perelomov's variables $\gamma =
-4\alpha^{2}\theta^{2}$ and $\nu=-(1+\alpha)^{2}/4\alpha$.
This completes Step~2 of our program.

Now we turn our attention to Step~3, and focus on Ito and
Fokker-Planck equations. 

The following idioms lead quickly to Theorem~\ref{th:theorem1}.
We define a
data three-vector $\lb{d}_{n} =
(d_{1,n},d_{2,n},d_{3,n})$ by \[ d_{i,n}=
\begin{cases}
  +1&\text{for\ }\ket{\psi_{n+1}} \propto 
  A^{\text{c}}_{i}\,\ket{\psi_{n}}\,,\\
  -1&\text{for\ }\ket{\psi_{n+1}} \propto 
  B^{\text{c}}_{i}\,\ket{\psi_{n}}\,.\\ 
\end{cases}
\]
Thus $\{\lb{d}_1,\lb{d}_2,\ldots\}$ is binary-valued data.
The mean
\begin{equation}
  E[\lb{d}_{n}]  =
  \braket{\psi_{n}}{A^{\text{c}\dagger}_{k}A^{\text{c}}_{k}}{\psi_{n}}
  -\braket{\psi_{n}}{B^{\text{c}\dagger}_{k}B^{\text{c}}_{k}}{\psi_{n}}
\end{equation}
satisfies
\begin{equation}
  \label{eq:measurement gain}
  E[\lb{d}_{n}] = 2\theta\,E[\lb{x}_{n}] \equiv
  g_{\text{s}}E[\lb{x}_{n}]
\end{equation}
which defines $g_{\text{s}} = 2\theta$ as the sensor gain.
We remark that $\lb{d}_{n}/g_{\text{s}}$ is therefore an
unbiased measure of $\lb{x}_{n}$.

We next define a zero-mean stochastic variable
$\lb{W}_{n}$ by
\begin{equation}
  \lb{d}_{n} =  g_{\text{s}}\,\lb{x}_{n} + \lb{W}_{n}\,,
\end{equation}
such that (to leading order in $\theta$) $\lb{W}_{n}$ has
the second-order stochastic properties of a discrete
Wiener increment:
\begin{equation}
E[(\lb{W}_{n})_{k}(\lb{W}_{n'})_{k'}] =
\delta_{nn'}\delta_{kk'}\,.
\end{equation}
An identity valid for $\ket{\psi_{n}}$ a coherent state,
\begin{equation}
  \label{eq:coherent identity}
  \braket{\psi_{n}}{s_{k}s_{l}}{\psi_{n}} = 
  \myonehalf j\delta_{kl} + j(j-\myonehalf) x_{k}x_{j}
  +\myonehalf i \epsilon_{klm}x_{m}\,,
\end{equation}
gives rise to an Ito increment of conventional form
\begin{equation}
  \label{eq:Ito}
  \delta\lb{x}_{n} = \lb{x}_{n+1}-\lb{x}_{n} 
  = g_{\text{s}}^{2}\lb{a}(\lb{x}_{n})  + g_{\text{s}}\,\lb{b}(\lb{x}_{n}) 
  \gsb\cdot \lb{W}_{n}\,,
\end{equation}
where $\lb{a}$ is called the drift vector and $\lb{b}$ is
called the diffusion matrix.

As a check, Lemma~\ref{th:triaxial feedback} requires that
the stochastic motion of $\lb{x}_{n}$ be confined to a
unit sphere, and so does the coherent state identity
\myeq{eq:coherent identity}, since it is inhomogenous in
$\lb{x}$. For consistency, therefore, the increment of the
$m$'th radial moment must vanish for all $m$ when
$|\lb{x}|=1$. This increment is readily calculated have
the general form
\begin{multline}
  \delta E_{n}[|\lb{x}|^{m}] 
  \propto
  \myonehalf m(m-2)[\lb{x}_{n}\gsb\cdot\lb{b}
  (\lb{x}_{n})\cdot\lb{b}^{\dagger}(\lb{x}_{n})\gsb\cdot\lb{x}_{n}]\,,\\
  + m|\lb{x}_{n}¥|^{2}
  [\myonehalf\mytr\lb{b}(\lb{x}_{n})\cdot\lb{b}^{\dagger}(\lb{x}_{n})+
  \lb{x}_{n}\gsb\cdot\lb{a}(\lb{x}_{n})]\,.
  \label{eq:on the sphere}
\end{multline}
Upon computing $\lb{a}(\lb{x})$ and $\lb{b}(\lb{x})$
explicitly we find
\begin{subequations}
\begin{align}
  \label{eq:a}
  \lb{a}(\lb{x}) =\ &\tfrac{1}{4}\lb{x}\, 
  [\alpha(1-2j)\lb{x}\gsb\cdot\lbhat{t}-
  (1+\myonehalf\alpha^{2})]\,\notag\\
  &+\,\tfrac{1}{4}\lbhat{t}\,[\,\alpha (1+2j)-
  \myonehalf\alpha^{2}\,\lb{x}\gsb\cdot\lbhat{t}\ ]\,,\\
  \lb{b}(\lb{x}) =\  & \tfrac{1}{2}[\lb{I}-\lb{x}\otimes\lb{x}+
  \alpha (\lbhat{t}\otimes\lb{x}-\lb{x}\gsb\cdot\lbhat{t}\,\lb{I})]\,,
  \label{eq:b}
\end{align}
\end{subequations}
such that the radial increment \myeq{eq:on the sphere}
indeed vanishes.

A Fokker-Planck equation for $P_{j}(\lbhat{x})$ is readily
found from (\ref{eq:Ito}) and (\ref{eq:a}--b). Setting
$z=\lbhat{x}\gsb\cdot\lbhat{t}$ we obtain
\begin{multline}
  0 = - \tfrac{\partial}{\partial z}
  [\alpha (1+z^{2}) + 2 j \alpha (1-z^{2}) -z(1+\alpha^{2})] P_{j}(z)\\
  +\tfrac{1}{2}\tfrac{\partial^{2}}{\partial z^{2}} 
  [(1-z^{2})(1-2\alpha z+\alpha^{2})]P_{j}(z)\,,
\end{multline}
which has the unique properly normalized solution
\begin{equation}
  P_{j}(\lbhat{x}) = \left[\frac{1-\alpha^{2}}{1-2 \alpha\,
  \lbhat{x}\gsb\cdot\lbhat{t}+\alpha^{2}}\right]^{2j+2}
  \hspace{-1em}.
\end{equation}
Substituting $\alpha = -\tanh\tfrac{1}{4}\beta$ per
\myeq{eq:alpha beta} yields Theorem~\ref{th:theorem1}.
This completes the third and final step of our proof

We now discuss some practical implications for experiments
in magnetic resonance force microscopy (MRFM) and gravity
wave (GW) interferometry. We first write the Ito equations
\myeq{eq:Ito} in Langevin form by substituting
\begin{subequations}
\begin{align}
\delta\lb{x}_{n} &\to \textstyle{\int_{0}^{t}\!dt'}\lbdot{x}(t')&
\lb{a}(\lb{x}_{n}) &\to r \textstyle{\int_{0}^{t}\!dt'}\lb{a}\lb(\lb{x}(t')\lb)\\
\label{eq:Langevin noise}
\lb{W}_{n} & \to rg_{\text{s}} \textstyle{\int_{0}^{t}\!dt'}\lb{x}^{\text{N}}(t')
\hspace{-0.25em}&
\lb{b}(\lb{x}_{n}) & \to \lb{b}\lb(\lb{x}(t)\lb)
\end{align}
\end{subequations}
where $\delta t=1/r$ is an interval and
$\lb{x}^{\text{N}}(t)$ is white noise
\begin{equation}
  \label{eq:white noise}
  E[x^{\text{N}}_{k}(t)x^{\text{N}}_{k'}(t')] =
\delta_{kk'}\delta(t-t')/g_{\text{s}}^{2}r\,.
\end{equation}
Then taking $\partial/\partial t$, the resulting Langevin
equation is
\begin{subequations}
\begin{equation}
  \label{eq:Langevin}
  \lbdot{x} = rg_{\text{s}}^{2}
  [\lb{a}(\lb{x})+ \lb{b}(\lb{x})\gsb\cdot(\lb{x}^{\text{M}}-\lb{x})]\,,
\end{equation}
where $\lb{x}^{\text{M}}(t) =
\lb{x}(t)+\lb{x}^{\text{N}}(t)$ is the measured spin axis.

We see that $\lb{x}(t)$ is dynamically attracted toward
the measured axis $\lb{x}^\text{M}(t)$. Even open-loop
spinometers exhibit this attraction, since for $\alpha=0$
we find
\begin{equation}
  \label{eq:Langevin II}
  \lbdot{x}|_{\alpha=0} = rg_{\text{s}}^{2}
  [-\tfrac{1}{4} \lb{x}+\tfrac{1}{2}(\lb{I}-\lb{x}\otimes\lb{x})
  \gsb\cdot(\lb{x}^{\text{M}}-\lb{x})]\,.
\end{equation}
\end{subequations}
\noindent Remark: a similar einselection-by-attraction is
evident even in uniaxial spinometry, where it dynamically
generates the asymptotic ``collapse'' of $\ket{\psi_{n}}$
to an eigenstate, as described by \myeq{eq:theoremI} and
Theorem~\ref{th:uniaxial}.

We now transform (\ref{eq:Langevin}--b) to the
second-order Newtonian equation of an oscillator. To do
this, we introduce a mass $m$ and frequency $\omega_{0}$
by defining
\begin{subequations}
\begin{align}
q^{\text{op}} &= (\hbar/j m\omega_{0})^{1/2} \
[+s_{2}\cos(\omega_{0}t)-s_{1}\sin(\omega_{0}t)]\,,\\ 
p^{\text{op}} &= (m\omega_{0}\hbar/j)^{1/2} \
[-s_{2}\sin(\omega_{0}t)-s_{1}\cos(\omega_{0}t)]\,.
\end{align}
\end{subequations}
We verify that for states with $z\sim 1$ the canonical
commutator $[q^{\text{op}},p^{\text{op}}] = i\hbar s_{3}/j
\simeq i\hbar$ holds. Defining the coherent oscillator
coordinate $q(t)$ to be
\begin{equation}
  q(t) = (j\hbar/ m\omega_{0})^{1/2} 
  (y(t)\cos\omega_{0}t-x(t)\sin\omega_{0}t)\,,
\end{equation}
we find that \myeq{eq:Langevin II} takes the equivalent
Newtonian form
\begin{align}
  \label{eq:Newtonian one}
  m\sddot{q}&= -m\omega_{0}^{2}\,q+f^{\text{ext}}+f^{\text{N}}
  & 
  q^{\text{M}} & =q+q^{\text{N}} \,,
\end{align}
where $f^{\text{ext}}(t)$ is an arbitrary external force. From
\myeq{eq:white noise} we obtain the noise spectral
densities, which satisfy
\begin{subequations}
\begin{align}
  \label{eq:SQL}
  S_{f^{\text{N}}\!f^{\text{N}}}(\omega)S_{q^{\text{N}}
  q^{\text{N}}}(\omega)& = \tfrac{1}{4} \hbar^{2} 
  \,,\\
\label{eq:Hilbert correlation}
S_{q^{\text{N}}\!f^{\text{N}}}(\omega)& =\tfrac{1}{2} 
i\hbar \mysgn\omega\,.
\end{align}
\end{subequations}
Here $S_{q^{\text{N}}f^{\text{N}}}(\omega) \equiv
\int_{-\infty}^{\infty}\!d\tau\,
e^{-i\omega\tau}E[q^{\text{N}}(t)f^{\text{N}}(t+\tau)]$.

We recognize \myeq{eq:SQL} as the standard quantum limit
(SQL), but the expression \myeq{eq:Hilbert correlation}
for $S_{q^{\text{N}}\!f^{\text{N}}}$ is surprising. We
will call it the \emph{Hilbert correlation} because it
asserts that $f^{\text{N}}(t)$ is proportional to the
Hilbert transform of $q^{\text{N}}(t)$. Physically, it
ensures that fluctuations in the observed position
$q^{\text{M}}(t)$ are accompanied by force fluctations
such that $q(t)$ is attracted toward $q^{\text{M}}(t)$, as
(\ref{eq:Langevin}--b) requires (and as is ubiquitious in
spinometry).

Is the Hilbert correlation observable? From
(\ref{eq:Newtonian one}) it can be shown that it is
undetectable in $q^{\text{M}}(t)$'s response to the
classical force $f^{\text{ext}}(t)$. Thus the standard
quantum limit for classical force signals, as detected in
experiments like GW interferometry, is not altered by the
presence (or absence) of a Hilbert correlation.

The author is presently investigating the conditions under
which Hilbert correlations \emph{are} observable; in
spinometric terms this seems to require a curved
(nonlinear) dynamical manifold, as can arise, \emph{e.g.}, 
in MRFM when several spins are present. 

A major obstacle is that the differential geometry and
topology of drift and diffusion functions on Hilbert
spaces are poorly understood. The following would help:
\begin{problem}\emph{Given a finite-dimensional Hilbert
space and on that space a set of Ito drift functions
$\lb{a}(\ket{\psi})$ and diffusion functions
$\lb{b}(\ket{\psi})$, either exhibit a set of spinometer
operators $\{A_{i},B_{i}\}$ that generate $\lb{a}$ and
$\lb{b}$, or prove that no such operators exist.}
\end{problem}

When students ask ``How does the Stern-Gerlach effect
work?'' or ``How does the standard quantum limit work?''
the author increasingly answers them in terms of drift and
diffusion functions. This focusses on the well-posed
problem of designing these functions and realizing them in
hardware, and yet makes clear how little is known, and how
much remains to be discovered.


\end{document}